\documentclass[5p, twocolumn]{elsarticle}

\usepackage{listings}
\usepackage{xcolor}

\usepackage{lineno,hyperref}
\modulolinenumbers[5]

\journal{Journal of \LaTeX\ Templates}

\usepackage{lipsum}

\usepackage{caption}

\usepackage{amsmath,graphicx}
\newlength{\fslength}
\makeatletter
\newcommand{\funnyP}{%
    \setlength{\fslength}{\f@size pt}%
    \reflectbox{${\mbox{P}}$}\hspace*{-.400\fslength}\mbox{${\mbox{P}}$}%
}
\newcommand{\funnyPbar}{%
    \setlength{\fslength}{\f@size pt}%
    \reflectbox{$\bar{\mbox{P}}$}\hspace*{-.400\fslength}\mbox{$\bar{\mbox{P}}$}%
}
\newcommand{\funnyPbarM}{%
    \setlength{\fslength}{\f@size pt}%
    \reflectbox{$\bar{\mbox{P}}$}\hspace*{-.400\fslength}\mbox{$\bar{\mbox{P}}_{M}$}%
}
\makeatother

\AtBeginDocument{%
  \providecommand\BibTeX{{%
    \normalfont B\kern-0.5em{\scshape i\kern-0.25em b}\kern-0.8em\TeX}}}

\usepackage{stackengine}
\usepackage{array}

\usepackage[flushleft]{threeparttable}
\usepackage{tabularx}

\begin{document}

\begin{frontmatter}

\title{Evaluating SYCL as a Unified Programming Model \\for Heterogeneous Systems}

\author{Ami Marowka }
\address{Parallel Research Lab, Israel }
\address{amimar2@yahoo.com}

%%%\author[mymainaddress,mysecondaryaddress]{Elsevier Inc}
%%%\ead[url]{www.elsevier.com}

%%%%\author[mysecondaryaddress]{Global Customer Service\corref{mycorrespondingauthor}}
%%%%%\cortext[mycorrespondingauthor]{Corresponding author}
%\ead{amimar2@yahoo.com}


\begin{abstract}

High-performance computing (HPC) applications are increasingly executed in heterogeneous environments, introducing new challenges for programming and software portability. SYCL has emerged as a leading model designed to simplify heterogeneous programming and make it more accessible to developers. Intended as a single-source, cross-platform parallel programming framework, SYCL promises portability, productivity, and performance across a variety of architectures.

However, these goals have not been consistently defined or realized, leaving developers with varying expectations. This paper addresses this gap by evaluating SYCL from the perspective of application developers. We analyze whether SYCL meets essential criteria for cross-platform development, including code portability, development productivity, and runtime efficiency.

Our evaluation draws on benchmarks and illustrative examples and focuses on SYCL's memory management and parallelism abstractions. We provide detailed comparisons between Unified Shared Memory (USM) and buffer-accessor approaches, as well as between NDRange and hierarchical kernel models. In addition to presenting our own benchmark results on Intel platforms, we synthesize findings from recent studies across multiple SYCL implementations and compilers.

Our results expose key limitations and inconsistencies in current SYCL implementations and offer insights into the steps needed to improve the framework's reliability and cross-platform usability.

\end{abstract}

\begin{keyword}
SYCL \sep Performance \sep Portability \sep Productivity \sep  heterogeneous computing
%\texttt{elsarticle.cls}\sep \LaTeX\sep Elsevier \sep template
%\MSC[2010] 00-01\sep  99-00
\end{keyword}

\end{frontmatter}

%\linenumbers

\section{Introduction}

The concept of deploying a single application binary across diverse environments has evolved significantly in software engineering---from virtual machines and containerization to the notion of packaging all required components into a single, portable executable \cite{Kurtzer}. In the context of heterogeneous programming, SYCL exemplifies this ambition by enabling developers to write unified applications that seamlessly target CPUs, GPUs, FPGAs, and other accelerators.

SYCL is an open standard developed by the Khronos Group \cite{Khronos} for single-source, cross-platform parallel programming in modern C++ \cite{SYCL}. It builds upon OpenCL and offers a range of kernel execution models, including single-instruction-multiple-threads (SIMT), data parallelism, and hierarchical parallelism. SYCL applications are compiled using a generalized backend architecture that supports specific targets such as OpenCL, CUDA, and HIP through intermediate representations (e.g., SPIR-V, PTX).

Though the first SYCL specification was introduced in 2014, it wasn't until around 2018 that practical implementations began to emerge. By 2022, SYCL development had gained significant traction, supported by major tools and vendors. Three leading implementations---Intel's DPC++ \cite{DPC}, Codeplay's ComputeCPP \cite{Codeplay}, and Heidelberg University's AdaptiveCpp \cite{Aksel, Alpay}---have helped shape the current ecosystem. Notably, Intel's acquisition of Codeplay has further consolidated its leadership in this space.

Despite SYCL's promise, its exact role and capabilities remain unclear to many developers. Official specifications emphasize SYCL's compilation strategies, including the single-source multiple-compiler pass (SMCP) and single-source single-compiler pass (SSCP) models. These determine whether separate or unified compiler passes are used for host and device code generation. However, these technical details often diverge from the practical concerns of developers, who seek ease of use, high productivity, code non-redundancy, functional portability, and consistent performance across platforms.

This paper aims to evaluate SYCL based on these expectations. While we acknowledge the growing interest in performance portability, we do not address it directly here. SYCL's current specification does not guarantee performance portability and explicitly states that it is not designed to enforce it. As such, we focus instead on three pillars of developer concern: portability, productivity, and performance.

Our study includes a detailed analysis of SYCL's memory management models (USM vs. buffer-accessors) and parallelism abstractions (NDRange vs. hierarchical kernels). We conduct benchmarks using Intel's tools and platforms, supplemented by a comprehensive review of related research. Our evaluation seeks to determine whether SYCL fulfills its promise as a practical and effective cross-platform programming model. We primarily focus on Intel's oneAPI DPC++ implementation, as it is currently the only vendor-supported and tightly integrated SYCL software-hardware stack, enabling a complete assessment against singularity requirements.

The main contributions of this paper are as follows:
\begin{itemize}
\item{We define core developer-centric requirements for a cross-platform programming model and evaluate SYCL against them.}

\item{We analyze how memory management strategies affect SYCL's portability and productivity.}

\item{We present benchmarks of SYCL kernels executed on Intel platforms using DPC++ and analyze the results.}

\item{We assess the performance consistency of SYCL's NDRange and hierarchical kernel models, as well as its USM and buffer abstractions.}

\item{We integrate findings from recent studies to provide a broader performance and portability assessment across multiple platforms and implementations.}
\end{itemize}

A preliminary version of this work appeared in the 34th Heterogeneity in Computing Workshop (HCW 2025) as part of the 39th IEEE International Parallel and Distributed Processing Symposium \cite{HCW}.
This article significantly extends our HCW 2025 work by adding new experiments, more abstraction coverage, and expanded analysis, as detailed in the contribution summary below:

\begin{itemize}
\item{The manuscript has been substantially rewritten for improved clarity, organization, and scientific depth.}
\item{Section 1 now includes additional background on SYCL, outlining its evolution and role in heterogeneous programming.}
\item{Sections 2 and 3 have been expanded with more detailed explanations and illustrative examples of SYCL abstractions.}
\item{Section 4 introduces two new code listings, demonstrating all combinations of data management and kernel models.}
\item{Section 6 adds four new benchmark evaluations, including novel results on AMD and NVIDIA GPUs to complement the Intel-based analysis.}
\item{Section 8 is entirely new, providing an in-depth Discussion of broader implications, recent specification changes, and challenges in achieving singularity.}
\item{Section 9 integrates four additional studies, strengthening the literature review and contextualizing our findings.}
\end{itemize}

The remainder of this paper is organized as follows:
Section 2 defines our criteria for evaluating SYCL and the motivation behind them.
Section 3 introduces the core abstractions in SYCL related to data and parallelism.
Section 4 presents illustrative examples.
Section 5 and Section 6 present benchmarking results.
Section 7 provides a comprehensive evaluation.
Section 8 discusses ongoing challenges in the SYCL ecosystem.
Section 9 reviews related work, and Section 10 concludes the paper.

\section{Singularity in Cross-Platform Programming}

This section introduces the concept of \textit{singularity} in the context of single-source, architecture-agnostic programming models such as SYCL. We define the term and present a set of non-functional requirements---performance, portability, and productivity---that a programming model must satisfy to be considered singular. These requirements serve as the foundation for our evaluation in subsequent sections.

\subsection{What Is Singularity?}

In heterogeneous programming, \textit{singularity} refers to the ability of a programming model to unify development across multiple hardware architectures, compilers, and backends through a single source codebase. A singular programming model allows developers to write code once and execute it consistently and efficiently across diverse platforms without modification, retuning, or loss of productivity.

In other words, singularity implies that the model abstracts away hardware-specific concerns while maintaining portability, productivity, and competitive performance.

\begin{quote}
\textbf{Definition --- Singularity:} \\
The singularity of a single-source, cross-platform programming model is its ability to satisfy the requirements of portability, productivity, and performance across diverse classes of hardware architectures, compiler implementations, and backend targets---without requiring changes to the source code.
\end{quote}

\subsection{Performance-Related Requirements}

From a singularity perspective, performance is not about achieving absolute optimality on every platform. Rather, it is about \textit{competitive consistency}: an application written in the model should perform reasonably well across platforms, and not degrade drastically when moved between implementations or updated toolchains.

Specifically, singularity requires:
\begin{itemize}
    \item The application's performance on one platform (e.g., CPU with OpenMP backend) should be comparable to native models (e.g., OpenMP).
    \item When ported to another platform (e.g., GPU with CUDA or HIP backend), the same code should perform comparably to manually optimized code written for that platform.
    \item Switching SYCL implementations (e.g., from DPC++ to AdaptiveCpp) or compiler versions should not result in performance drops of an order of magnitude or more.
    \item Interchangeable abstractions (e.g., NDRange vs. hierarchical kernels, or USM vs. buffer-accessors) should yield similar performance on the same device.
\end{itemize}

Singularity is broken if the choice of abstraction leads to significantly different runtime behavior on the same platform, forcing developers to rewrite or retune their code to maintain efficiency.

\subsection{Portability-Related Requirements}

Portability lies at the heart of singularity. It demands that developers be able to write code once and run it anywhere---across platforms, architectures, and SYCL implementations---without modifying the source.

Functional portability under singularity includes:
\begin{itemize}
    \item Correct compilation and execution of the same source code on CPUs, GPUs, and other accelerators.
    \item Cross-compatibility between different SYCL toolchains (e.g., DPC++, AdaptiveCpp) without requiring changes to the code structure or semantics.
    \item Stable behavior and portable performance across backend configurations (e.g., OpenCL, Level Zero, CUDA).
    \item Independence from vendor-specific compiler directives, memory alignment hacks, or hardware-bound tuning parameters.
\end{itemize}

If switching a compiler or backend requires code changes to restore functionality or performance, singularity is not achieved.

\subsection{Productivity-Related Requirements}

Singularity also implies a high level of \textit{developer productivity}. This does not merely mean shorter code or fewer lines---it means a development experience that minimizes friction, learning curves, and manual tuning across platforms.

A singular model must offer:
\begin{itemize}
    \item Simple abstractions that do not require deep knowledge of backend-specific behavior.
    \item Automatic memory management or portable interfaces (e.g., USM, buffer-accessors) that do not force developers to test and choose manually based on performance.
    \item Unified kernel abstractions (e.g., NDRange and hierarchical models) that offer similar results without platform-specific adjustments.
\end{itemize}

Singularity is diminished if developers are required to run benchmarks to determine which abstraction performs better, or if productivity hinges on undocumented runtime behaviors (such as when buffers are implicitly copied).

\subsection{Summary}

Singularity represents a convergence of three key attributes in a programming model: \textbf{portability}, \textbf{productivity}, and \textbf{performance}. A model that achieves singularity abstracts platform-specific concerns without sacrificing runtime efficiency or developer usability.

For example, a matrix multiplication SYCL kernel should:
\begin{itemize}
    \item Run correctly and efficiently on both CPU and GPU without any source changes.
    \item Deliver performance within an acceptable range of native implementations like OpenMP or CUDA.
    \item Allow developers to write and maintain the kernel without platform-specific logic or manual tuning.
\end{itemize}

In the sections that follow, we assess the degree to which SYCL achieves singularity, using both benchmarks and qualitative analysis of its core abstractions and implementations.

\section{Data Management and Parallelism Abstractions of SYCL}

To understand how SYCL approaches singularity, it is essential to examine the core abstractions it offers for memory management and parallelism. These abstractions are central to application performance and portability, and significantly influence the developer's productivity. In this section, we briefly describe the key features of four SYCL abstractions that are directly relevant to our evaluation.

\subsection{Data Management Abstractions}

SYCL provides two primary mechanisms for managing data transfers between the host and the device:

\begin{itemize}
  \item \textbf{Unified Shared Memory (USM)}
  \item \textbf{Buffer-accessor model}
\end{itemize}

These represent two fundamentally different paradigms in memory management.

\smallskip
\smallskip
\textbf{Unified Shared Memory (USM):} \\
USM is a pointer-based approach that creates a unified address space accessible by both the host and device. It relies on hardware and runtime support to maintain data consistency and to migrate data between memory spaces automatically. This significantly reduces the burden on the programmer, who no longer needs to manually manage memory movement or synchronization.

USM supports implicit memory management, freeing developers from tracking where updated data reside or orchestrating manual data transfers. This model offers high productivity and closely resembles conventional pointer-based programming in C++.
While more optimized USM handling is possible, the examples in the next section reflect portable usage without hardware-specific tuning, in line with singularity requirements.

\smallskip
\smallskip
\textbf{Buffer-accessor model:} \\
In contrast, the buffer-accessor approach uses explicit abstractions to manage data. Buffers are containers that represent data objects, and access to these buffers is performed via accessor objects. The accessor model informs the runtime about how data will be used and enables partial or implicit memory transfers.

Unlike USM, the buffer-accessor model requires the programmer to be aware of the runtime's data synchronization behavior. For example, developers may need to define whether data should be transferred back to the host after kernel execution using methods like \texttt{set\_final\_data()} or \texttt{set\_write\_back()}. While this provides more control, it also introduces complexity and potential pitfalls.

\smallskip
\smallskip
\textbf{Comparison and Implications:} \\
The USM model is more implicit and user-friendly but requires runtime and hardware support. The buffer-accessor model is more explicit and portable but can reduce productivity due to its verbosity and the need for fine-grained runtime awareness.
While the programmer can influence when data is moved through buffer lifetime and accessor scope, the SYCL runtime is ultimately responsible for performing the actual data transfers.

A singular programming model would ideally allow either abstraction to be used interchangeably---without significantly affecting performance or behavior. However, as shown later in our benchmarks, this is not yet the case with SYCL.

\subsection{Parallelism Abstractions}

SYCL supports four kernel execution models for expressing parallelism, two of which are particularly relevant to our analysis:

\begin{itemize}
  \item \textbf{NDRange kernels}
  \item \textbf{Hierarchical parallel kernels}
\end{itemize}

These abstractions map SYCL kernels to hardware in different ways, often influencing performance and portability.

\smallskip
\smallskip
\textbf{NDRange Kernels:} \\
The NDRange model is inspired by GPU-style parallelism, such as CUDA or OpenCL. It defines a flat hierarchy of parallel execution, where global and local work-item IDs are explicitly specified. Programmers are responsible for manually managing synchronization using barriers.

This model is suited to fine-grained, massively parallel architectures like GPUs, and gives developers precise control over parallel execution.

\smallskip
\smallskip
\textbf{Hierarchical Kernels:} \\
Hierarchical kernels are inspired by nested parallelism models such as OpenMP. They support multiple levels of parallel execution, including work-groups and work-items, and introduce implicit synchronization between levels. This abstraction is naturally suited to CPU architectures with coarse-grained parallelism.

Hierarchical kernels abstract away some of the lower-level synchronization logic and map cleanly to OpenMP backends.

\smallskip
\smallskip
\textbf{Comparison and Trade-offs:} \\
Although NDRange and hierarchical kernels share a two-level structure (e.g., group and work-item), they differ significantly in implementation. NDRange requires explicit barriers, while hierarchical kernels perform synchronization implicitly.

In theory, a singular model would ensure that these two abstractions are interchangeable in terms of performance and portability. In practice, as shown in Section~6, the performance of these models varies considerably depending on the backend and device, thereby undermining singularity.

\smallskip
\smallskip
\textbf{Summary:} \\
SYCL's abstraction layers are designed to offer flexibility, but that flexibility comes with trade-offs. The choice between USM and buffer-accessors, or between NDRange and hierarchical kernels, has significant implications for portability, productivity, and performance. The lack of consistency among these abstractions across devices and implementations presents a challenge to SYCL's singularity.

\section{Illustrative Examples} 

To better understand how SYCL's abstractions influence portability, productivity, and performance---and thus its singularity---we present four representative implementations of a reduction algorithm. These examples illustrate all combinations of SYCL's two main memory management models (USM and buffers) and two primary kernel types (NDRange and hierarchical). Each version performs a basic reduction over a vector of integers. The differences in code structure, abstraction usage, and execution semantics highlight the trade-offs SYCL imposes on developers when choosing among these options.

\definecolor{codegray}{rgb}{0.5,0.5,0.5}
\begin{lstlisting}[caption=Reduction using NDRange Parallel Kernel alongside Buffers, label={lst:listing-cpp}, language=C++, numbers=left,  basicstyle=\ttfamily\scriptsize, keywordstyle=\color{blue},
numberstyle=\tiny\color{codegray},xleftmargin=.025\textwidth]
#include <sycl/sycl.hpp>
using namespace sycl;

auto Reduction_ND_Buffer_kernel
        (queue &q, vector<int> &V, size_t N, size_t M)  
{{
buffer BUFa(V);
std::vector<int> total(1);
total[0] = 0;
buffer<int> bufSum {total}; // reduction result

auto e = q.submit([&](handler &h){
 accessor A(BUFa, h, read_only);    
 local_accessor<int> Local_V{M, h};   
 auto sum = bufSum.get_access<access::mode::atomic>(h);	

 h.parallel_for
     (nd_range<1>{{N},{M}},[=](nd_item<1> item){
  const int iG = item.get_global_id(0);
  const int jL = item.get_local_id(0);    
  Local_V[jL] = A[iG];          
  item.barrier(access::fence_space::local_space);
  if(item.get_group().leader()){
     int tmp = 0;
     for(int j = 0; j < M; ++ j)
        tmp += Local_V[j];
     sum[0].fetch_add(tmp); // Atomic		
    }
  });
 });
};
std::cout << "[0]=" << total[0] << "\n";         
return 0;  
}
\end{lstlisting}     
%frame=single,

We detail four versions of reduction-based applications of SYCL. The first version demonstrates the use of an NDRange parallel kernel alongside a buffer abstraction (Listing 1), 
the second version of reduction uses a hierarchical parallel kernel alongside a USM abstraction (Listing 2)
the third version of reduction uses an NDRange parallel kernel alongside a USM abstraction (Listing 3)
and the fourth version of reduction uses a hierarchical parallel kernel alongside a buffer abstraction (Listing 4). 
These examples, in conjunction with past studies that have used various benchmark applications, will allow us to analyze the impact of these abstractions on the singularity-related requirements of performance, productivity, and portability of SYCL to better assess its degree of singularity.

\subsection{NDRange Kernel and Buffer}

Listing 1 shows the version of SYCL reduction that uses the NDRange parallel kernel and buffer data objects. Four parameters are passed to the reduction function (lines 4-5): A queue {\it q} is defined, and has an associated default device for offloading the kernel from the host to the device (not shown). A vector {\it V} of size {\it N} contains the data, where the size of the local memory work group is {\it M}. A buffer {\it BUFa} is constructed and initialized by the data of the input vector {\it V}, and an array container {\it total} is defined to store the results of reduction (lines 8-9). Its data are copied into a buffer {\it bufSum} during its construction (line 10). Following this, three accessor objects are created: an accessor {\it A} for buffer {\it BUFa} with access tag {\it read\_only} (line 13), a local accessor {\it Local\_V} for accessing the local memory, and an atomic accessor {\it sum} for the buffer {\it bufSum}.

The NDRange parallel reduction kernel is now ready to be launched on the device (lines 17-29). The global and local indices of each item are first assigned to variables {\it iG} and {\it jL}, respectively (lines 19 and 20). The data are then copied from the global memory to the local memory (line 21), and this is followed by the execution of a barrier synchronization operation that handles waiting for the copying of the data from the global memory (line 22). Once all the data have been copied to the local memory, the local reduction operation is initiated by the leader of the group (lines 24-26). The result of local reduction is stored in a {\it tmp} variable, and is then added atomically to {\it bufSum} by using the {\it sum} accessor (line 27). The {\it total} result of reduction stored in {\it bufSum} is then implicitly transferred from the device to the host (line 31) so that it can be displayed to the user (line 32). 

It is worth noting that in this version of reduction, the data are implicitly transferred from the host to the device, and vice versa. {\it BUFa} is transferred from the host to the device on line 21, immediately after the first reference is made by its accessor {\it A} to read the data from the global memory. {\it bufSum} is transferred from the host to the device on line 27, immediately after the first reference is made by its accessor {\it sum} to add the result of local reduction stored in {\it tmp}.

Once the kernel has been executed, the final result of reduction stored in {\it bufSum} on the device is transferred back to the host before {\it bufSum} is destroyed, and leave its scope in line 31. Because {\it bufSum} was created and initialized from the host array {\it total}, the latter is updated with the final result of reduction when {\it bufSum} is destroyed.

\subsection{Hierarchical Kernel and USM}

Listing 2 shows the version of SYCL reduction that uses the hierarchical parallel kernel and the USM. 
First, the USMs are allocated: one for the input {\it data} (line 4) and one for the {\it total} variable that stores the final result of reduction (line 7). These USMs can be accessed on the host and the device, and are initialized after they have been allocated (lines 5 and 8). Following this, the local accessor {\it Local\_V} is defined for accessing the local memory (line 11) and the number of groups, {\it NG}, is calculated (lines 12 and 13).

The hierarchical parallel reduction kernel is now ready to be launched on the device (lines 16-36). The atomic variable {\it total\_atomic} is first defined to allow atomic operations on the {\it total} shared memory variable (lines 18-21). The global and local indices of each item are then assigned to the variables {\it iG} and {\it jL}, respectively (lines 24 and 25). The {\it data} are then copied from the global memory to the local memory (line 26), and this is followed by an {\bf implicit} barrier synchronization operation that handles the wait for the copying of {\it data} from the global memory (line 27). Once all the {\it data} have been copied to the local memory, the local reduction operation is initiated by the leader of the group (lines 29-32). The result of local reduction is stored in a {\it tmp} variable and then added atomically to the {\it total} variable by using the {\it total\_atomic} variable (line 33). The final result of reduction stored in {\it total} is then transferred implicitly from the device to the host (line 37) so that it can be displayed to the user (line 37).

\definecolor{codegray}{rgb}{0.5,0.5,0.5}
\begin{lstlisting}[caption=Reduction using Hierarchical Parallel Kernel alongside Unified Shared Memory, label={lst:listing-cpp}, language=C++, numbers=left,  basicstyle=\ttfamily\scriptsize, keywordstyle=\color{blue},
numberstyle=\tiny\color{codegray},xleftmargin=.025\textwidth]
auto Reduction_HR_USM_kernel
       (queue &q, vector<int> &V, size_t N, size_t M) 
{    
auto data = malloc_shared<int>(N, q);
for (int i = 0; i < N; i++) data[i] = V[i];
  
auto total = malloc_shared<int>(1, q);
total[0] = 0;

auto e = q.submit([&](handler &h){
 local_accessor<float> Local_V{M, h};    
 size_t NG = N/M;
 range num_groups{NG};         
 range group_size{M};
  
 h.parallel_for_work_group
    (num_groups, [=](group<1> grp){
      auto total_atomic = atomic_ref<int, 
      memory_order::relaxed, 
      memory_scope::device, 
      access::address_space::global_space>(total[0]);			
    grp.parallel_for_work_item
                (group_size,[&](h_item<1> item){
     const int iG = item.get_global_id(0);
     const int jL = item.get_local_id(0); 	            	
     Local_V[jL] = data[iG];
    });  // Implicit barrier
	
    if(grp.leader()){
       int tmp = 0;
       for(int j = 0; j < M; ++ j)
           tmp += Local_V[j];
       total_atomic += tmp; // Atomic						
    }
  });		
 });	
std::cout << "[0]=" << total[0] << "\n";  
free(data, q);         
return 0;  
}
\end{lstlisting} 
%frame=single,

It is worth noting that the {\it data} and {\it total} arrays are implicitly transferred from the host to the device, and vice versa. This transfer is carried out (lines 26 and 33) immediately after the first reference is made to read from the {\it data} and {\it total} arrays (by using {\it total\_atomic}). 
Once the kernel has been executed, the {\it total} array is transferred from the device back to the host immediately after the first reference is made to read the final result of reduction and display it to the user (line 37).

To complete the possible combinations of data management abstractions and parallelism abstractions for implementing Reduction applications, we present the combinations of NDRange parallel kernel and USM (Listing 3) and hierarchical parallel kernel and Buffers (Listing 4).

These examples underscore the following challenges to singularity:

\begin{itemize}
    \item \textbf{Productivity Gaps:} The buffer-accessor model introduces complexity that reduces ease of use. Developers must understand the accessor lifecycle and manage host-device transitions explicitly.
    \item \textbf{Portability Limits:} While all versions use SYCL, subtle behavioral differences (especially around memory synchronization and kernel execution) mean that consistent performance cannot be assumed across platforms.
    \item \textbf{Performance Variability:} As our benchmarks later show, the performance of these abstractions varies significantly across devices and SYCL implementations-violating the performance-related expectations of singularity.
\end{itemize}

\definecolor{codegray}{rgb}{0.5,0.5,0.5}
\begin{lstlisting}[caption=Reduction using NDRange Parallel Kernel alongside Unified Shared Memory, label={lst:listing-cpp}, language=C++, numbers=left,  basicstyle=\ttfamily\scriptsize, keywordstyle=\color{blue},
numberstyle=\tiny\color{codegray},xleftmargin=.025\textwidth]
#include <sycl/sycl.hpp>
using namespace sycl;

auto Reduction_ND_USM_kernel
        (queue &q, vector<int> &V, size_t N, size_t M)  
{
  auto data = malloc_shared<int>(N, q);
  for (int i = 0; i < N; i++) data[i] = V[i];
  
  auto total = malloc_shared<int>(1, q);
  total[0] = 0;

auto e = q.submit([&](handler &h){   
 local_accessor<int> Local_V{M, h};   	

 h.parallel_for
     (nd_range<1>{{N},{M}},[=](nd_item<1> item){
		 
  auto total_atomic = atomic_ref<int, 
      memory_order::relaxed, 
      memory_scope::device, 
      access::address_space::global_space>(total[0]);
  const int iG = item.get_global_id(0);
  const int jL = item.get_local_id(0);    
  Local_V[jL] = A[iG];          
  item.barrier(access::fence_space::local_space);
  if(item.get_group().leader()){
     int tmp = 0;
     for(int j = 0; j < M; ++ j)
        tmp += Local_V[j];
     total_atomic += tmp; // Atomic		
    }
  });
 });
std::cout << "[0]=" << total[0] << "\n"; 
free(data, q);        
return 0;  
}
\end{lstlisting} 
%frame=single,

In theory, a truly singular programming model would allow developers to choose any of these abstraction combinations and achieve comparable performance and behavior across platforms. In practice, the choice often depends on the target architecture and compiler behavior, which undermines that goal.

We now focus on benchmarking the impacts of the buffer and USM abstractions of data management as well as the NDRange and hierarchical parallel kernels with regard to the performance of SYCL applications. This is done based on the applications of reduction described above as well as other benchmarks used in past work.  

\bigskip

\definecolor{codegray}{rgb}{0.5,0.5,0.5}
\begin{lstlisting}[caption=Reduction using Hierarchical Parallel Kernel alongside Buffers, label={lst:listing-cpp}, language=C++, numbers=left,  basicstyle=\ttfamily\scriptsize, keywordstyle=\color{blue},
numberstyle=\tiny\color{codegray},xleftmargin=.025\textwidth]
auto Reduction_HR_Buffer_kernel
       (queue &q, vector<int> &V, size_t N, size_t M) 
{    
buffer BUFa(V);
vector<int> total(1);
std::total[0] = 0;
buffer<int> bufSum {total}; // reduction result

auto e = q.submit([&](handler &h){
 accessor A(BUFa, h, read_only);
 local_accessor<float> Local_V{M, h};    
 auto sum = bufSum.get_access<access::mode::atomic>(h);	
 size_t NG = N/M;
 range num_groups{NG};         
 range group_size{M};
  
 h.parallel_for_work_group
                (num_groups, [=](group<1> grp){
    grp.parallel_for_work_item
                (group_size,[&](h_item<1> item){
     const int iG = item.get_global_id(0);
     const int jL = item.get_local_id(0); 	            	
     Local_V[jL] = A[iG];
    });  // Implicit barrier
	
    if(grp.leader()){
       int tmp = 0;
       for(int j = 0; j < M; ++ j)
           tmp += Local_V[j];
       sum[0].fetch_add(tmp); // Atomic						
    }
  });		
 });	
host_accessor hc(bufSum, read_only);
std::cout << "[0]=" << total[0] << "\n";         
return 0;  
}
\end{lstlisting}

\begin{figure*} %[!htb]
    %\captionsetup{justification=centering} %,margin=2cm}
    %\caption{Comparing results of matrix multiplication application \\on the CPU and the GPU between Buffer and USM approaches.}\label{Fig:Data1}
\begin{minipage}{.49\linewidth} 
%\begin{subfigure}{0.49\linewidth} %[h] 
\caption{Reduction using Local Memory and NDRange kernel on Intel Max 1100 GPU and Xeon 8480 CPU.}     
     \centering
     \includegraphics [height=7cm, width=1.0\linewidth]{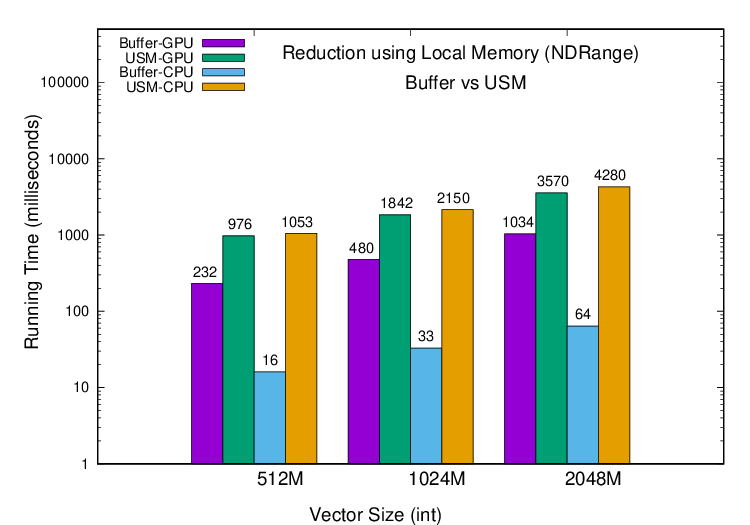}
\label{fig:fig11} 
\end{minipage} 
%\end{subfigure}
%\end{figure*} 
\hfill
%\begin{figure*}[!htb]
\begin{minipage}{.49\linewidth}
\caption{Reduction using Local Memory and hierarchical kernel on Intel Max 1100 GPU and Xeon 8480 CPU.}     
     \centering
     \includegraphics [height=7cm, width=1.0\linewidth]{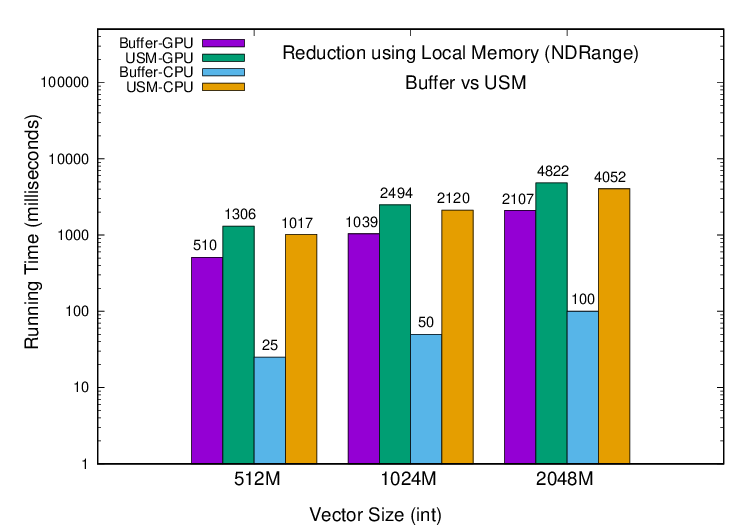}
\label{fig:fig12}
\end{minipage}
%\end{subfigure}
\end{figure*}

\section{Benchmarks: Buffer vs USM}

To evaluate SYCL's memory management abstractions from the perspective of singularity, we conducted a series of benchmarks comparing the performance of applications implemented using Unified Shared Memory (USM) and the buffer-accessor model. Our goal is to assess how these abstractions affect performance consistency and portability across architectures, and whether they can be used interchangeably without violating singularity.

\subsection{Benchmark Setup}

We performed our benchmarks on Intel hardware using the Intel Developer Cloud platform. This environment was selected because Intel is currently the only commercial vendor offering a complete SYCL toolchain.

\begin{itemize}
    \item \textbf{Device Platform:} Intel Data Center Max 1100 GPU
    \begin{itemize}
        \item 56 Xe-cores @ 1.55 GHz
        \item Peak memory bandwidth: 1228.8 GB/s
        \item Driver: 1.3.26516
        \item Backends: OpenCL Graphics, Intel Level Zero
    \end{itemize}
    \item \textbf{Device Platform:} NVIDIA A100 GPU
    \begin{itemize}
        \item 6912 CUDA cores @ 1.512 GHz
        \item Peak memory bandwidth: 1935 GB/s
        \item Driver: Codeplay 515.65.01
        \item Backends: CUDA 11.7 toolchain/runtime backend
    \end{itemize}
    \item \textbf{Device Platform:} AMD MI250 GPU
    \begin{itemize}
        \item 208 Compute Units @ 1.6 GHz
        \item Peak memory bandwidth: 3200 GB/s
        \item Driver: 30.10.1
        \item Backends: ROCm 7.1.0 HIP/AMDGPU backend
    \end{itemize}
    \item \textbf{Host Platform:} Intel Xeon Platinum 8480 CPU
    \begin{itemize}
        \item 56 cores @ 2 GHz
        \item 105 MB cache, 4 TB max memory
        \item Driver: 2023.16.7.0.21160000
    \end{itemize}
    \item \textbf{Software Stack:}
    \begin{itemize}
        \item SYCL 2020 standard
        \item Intel oneAPI DPC++ Compiler (2023.2.0)
        \item NVIDIA CUDA 11.7 
        \item AMD ROCm 7.1.0
    \end{itemize}
\end{itemize}

\begin{figure*} %[!htb]
    %\captionsetup{justification=centering} %,margin=2cm}
    %\caption{Comparing results of matrix multiplication application \\on the CPU and the GPU between Buffer and USM approaches.}\label{Fig:Data1}
\begin{minipage}{.49\linewidth} 
%\begin{subfigure}{0.49\linewidth} %[h] 
\caption{Comparing Results of Vector-Addition on the Intel GPU between Buffer and USM approaches using Level-Zero.}     
     \centering
     \includegraphics [height=7cm, width=1.0\linewidth]{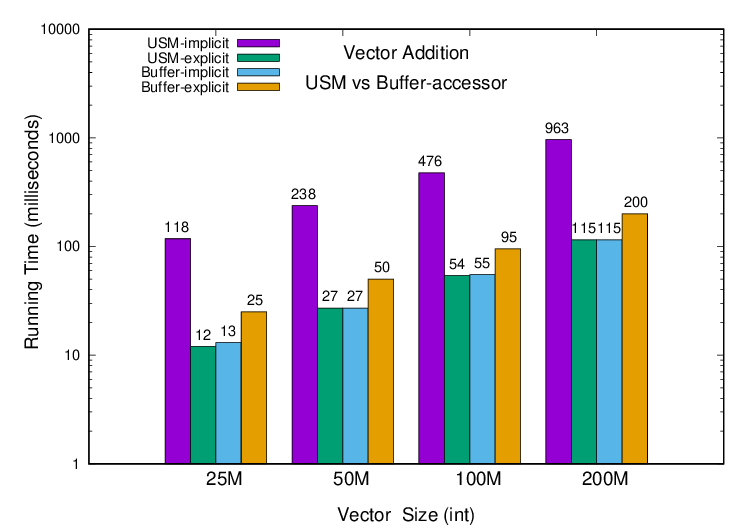}
\label{fig:fig22} 
\end{minipage} 
%\end{subfigure}
%\end{figure*} 
\hfill
%\begin{figure*}[!htb]
\begin{minipage}{.49\linewidth}
\caption{Comparing Results of Matrix Multiplication on the Intel CPU and the GPU between Buffer and USM approaches.}     
     \centering
     \includegraphics [height=7cm, width=1.0\linewidth]{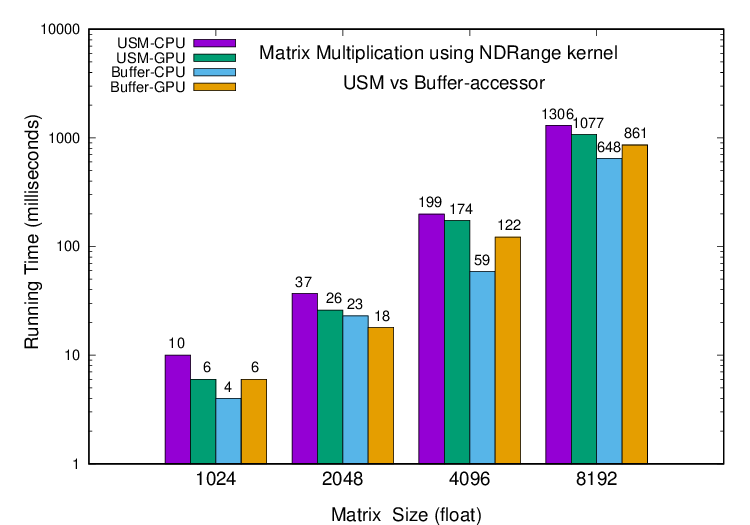}
\label{fig:fig44}
\end{minipage}
%\end{subfigure}
\end{figure*}

\begin{figure*} %[!htb]
    %\captionsetup{justification=centering} %,margin=2cm}
    %\caption{Comparing results of matrix multiplication application \\on the CPU and the GPU between Buffer and USM approaches.}\label{Fig:Data1}
\begin{minipage}{.49\linewidth} 
%\begin{subfigure}{0.49\linewidth} %[h] 
\caption{Comparing Results of Matrix Multiplication on the NVIDIA GPU between Buffer and USM approaches.}     
     \centering
     \includegraphics [height=7cm, width=1.0\linewidth]{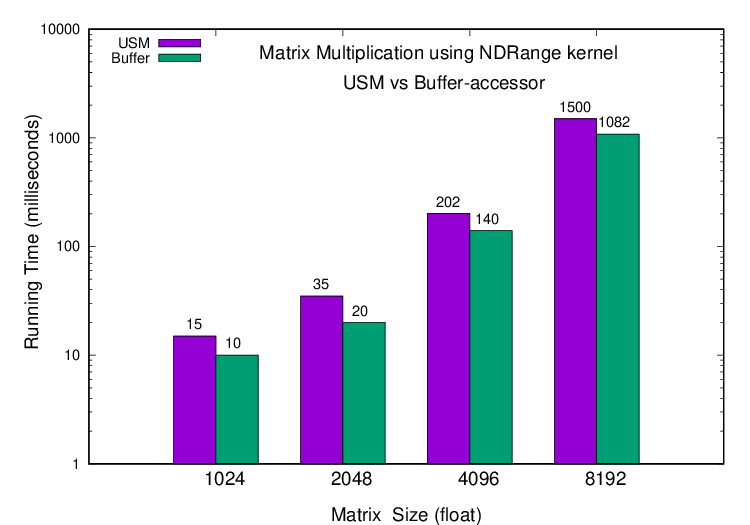}
\label{fig:fig22} 
\end{minipage} 
%\end{subfigure}
%\end{figure*} 
\hfill
%\begin{figure*}[!htb]
\begin{minipage}{.49\linewidth}
\caption{Comparing Results of Matrix Multiplication on the AMD GPU between Buffer and USM approaches.}     
     \centering
     \includegraphics [height=7cm, width=1.0\linewidth]{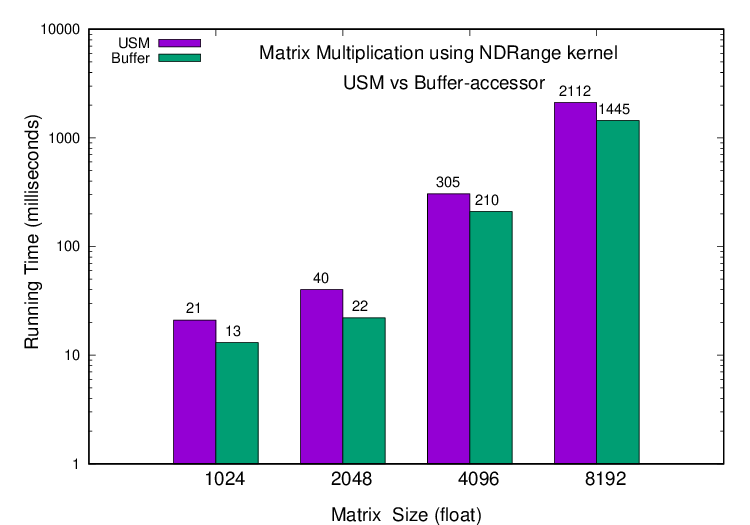}
\label{fig:fig44}
\end{minipage}
%\end{subfigure}
\end{figure*}

\subsection{Benchmark Results: Reduction Kernels}

Figures 1 and 2 summarize the performance (in milliseconds) of the reduction application using both NDRange and hierarchical kernels, with buffer-accessor and USM memory models, respectively.
Each measurement is an average of 50 runs, after a warm-up of 10 initial runs, for inputs of sizes 512 MB, 1024 MB, and 2048 MB. The aim of these benchmarks is to compare the performance of the USM and buffer-accessor approaches on the CPU and GPU when using a level-zero backend. 

\smallskip
\smallskip
\textbf{Key observations:}
\begin{itemize}
    \item Across all input sizes, the buffer-accessor model consistently outperformed the USM model on both the CPU and GPU.
    \item The performance gap was most pronounced on the CPU, where buffer-accessors outpaced USM by up to \textbf{66.9$\times$} using NDRange kernels and up to \textbf{42.4$\times$} using hierarchical kernels.
    \item On the GPU, buffer-accessors maintained a lead over USM by margins of up to \textbf{4.2$\times$} (NDRange) and \textbf{2.6$\times$} (hierarchical).
\end{itemize}

These results clearly demonstrate that the two memory models are not interchangeable in performance terms---a violation of singularity's performance criterion.

\subsection{Vector Addition: Explicit vs. Implicit Transfers}

We also benchmarked a vector addition application to examine the effect of explicit versus implicit memory transfers under both memory models.

\begin{itemize}
    \item \textit{Implicit transfers} allow the runtime to automatically handle data movement.
    \item \textit{Explicit transfers} require the developer to manually orchestrate data migration.
\end{itemize}

Figure~3 shows that:
\begin{itemize}
    \item USM with implicit transfers performed the worst, trailing by up to \textbf{9.8$\times$} compared to other modes.
    \item The best results were achieved with \textbf{USM using explicit data migration}, matching the performance of buffer-accessors with explicit transfers.
    \item Implicit buffer-accessor behavior delivered similar performance to its explicit counterpart, showcasing stronger portability. Moreover, it shows that the runtime can automatically optimize implicit buffer-accessor transfers (e.g., overlap of communication and computation), whereas explicit synchronization may introduce additional overhead.
\end{itemize}

These results highlight a key problem: the performance of USM varies drastically depending on whether memory movement is implicit or explicit, while buffer-accessors remain relatively stable.

\subsection{Matrix Multiplication: NDRange Comparison}

To assess how the two memory models scale with compute intensity, we benchmarked a tiled matrix multiplication kernel using NDRange on both Intel CPU and GPU (Figure~4) and NVIDIA and AMD GPUs.
Input matrix sizes ranged from 1024 to 8192 (float elements).

\smallskip
\smallskip
\textbf{Findings:}
\begin{itemize}
    \item Buffer-accessors consistently outperformed USM across all matrix sizes.
    \item On the CPU, the buffer-accessor model was up to \textbf{3.4$\times$} faster.
    \item On the GPU, the performance advantage was smaller but consistent (up to \textbf{1.81$\times$}).
\end{itemize}

This result reinforces the pattern observed in reduction and vector addition: buffer-accessors yield more predictable and superior performance across architectures.

\subsection{Discussion}

These benchmarks reveal the following insights relevant to singularity:

\begin{itemize}
    \item \textbf{USM is not performance-portable.} Its overhead in implicit mode is significant, particularly on CPUs, where memory migration costs dominate execution time.
    \item \textbf{Buffer-accessors are more stable.} While more complex in syntax, they provide consistent and predictable performance across devices and compilers.
    \item \textbf{Manual tuning is required.} Developers must decide between implicit and explicit strategies, and between memory models, based on trial-and-error and platform knowledge. This undermines both productivity and portability.
\end{itemize}

%%\begin{figure*}[!htb]
    %\captionsetup{justification=centering} %,margin=2cm}
%%         \caption{Comparing results of vector-addition on the GPU between Buffer and USM approaches 
%%				using Level-Zero.}\label{Fig:Data1}
%%     \centering
%%     \includegraphics [height=7cm, width=0.6\linewidth]{fig2.eps}

%%\label{fig:fig2}  
%%\end{figure*}   

\begin{figure*} %[!htb]
    %\captionsetup{justification=centering} %,margin=2cm}
    %\caption{Comparing results of matrix multiplication application \\on the CPU and the GPU between Buffer and USM approaches.}\label{Fig:Data1}
\begin{minipage}{.49\linewidth} 
%\begin{subfigure}{0.49\linewidth} %[h] 
\caption{Comparing results of Reduction using Local Memory and Buffers on the CPU and the GPU between NDRange and hierarchical Parallel Kernels.}     
     \centering
     \includegraphics [height=7cm, width=1.0\linewidth]{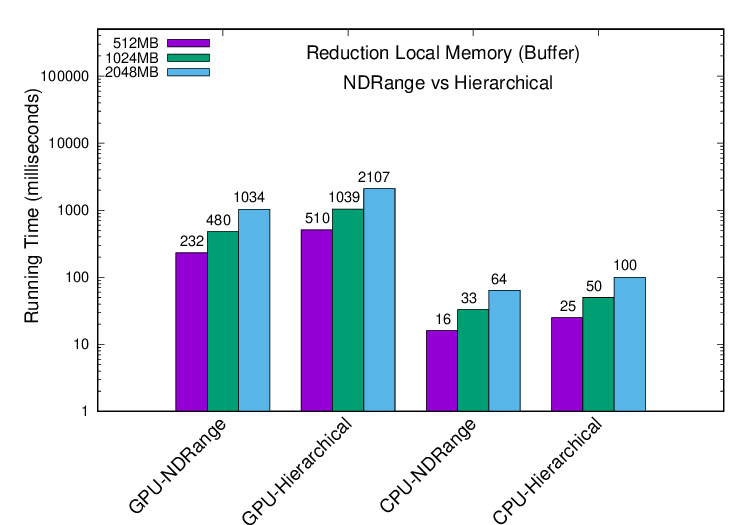}
\label{fig:fig9} 
\end{minipage} 
%\end{subfigure}
%\end{figure*} 
\hfill
%\begin{figure*}[!htb]
\begin{minipage}{.49\linewidth}
\caption{Comparing results of Reduction using Local Memory and USM on the CPU and the GPU between NDRange and hierarchical Parallel Kernels.}     
     \centering
     \includegraphics [height=7cm, width=1.0\linewidth]{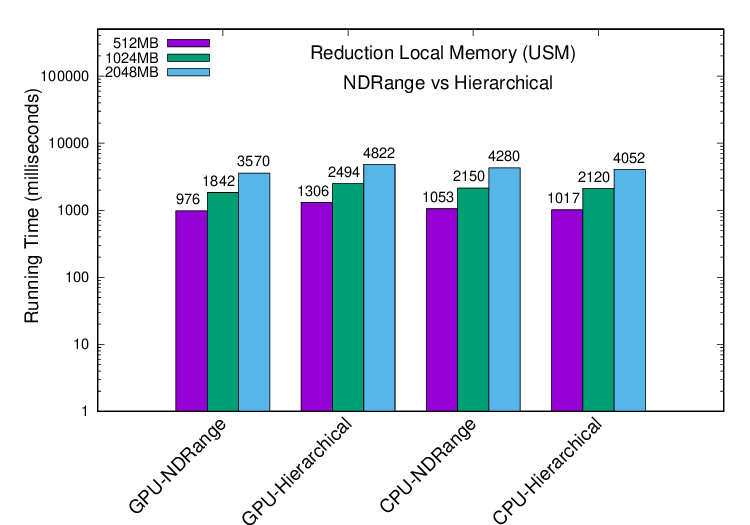}
\label{fig:fig10}
\end{minipage}
%\end{subfigure}
\end{figure*}

Are these results consistent with other studies? The answer is a resounding yes. Studies from recent years \cite{Nozal,Vetter,Joube,Gu,Mijic,Marcel,Lin, 
Chien, Czarnul, Fumero}
have clearly shown that the USM-based approach yields poor performance when using {\it fully implicit} data migration. For example, Jin and Vetter \cite{Vetter} evaluated the performance of a kernel for vector addition that used UM supported by AMD's HIP API against non-UM versions in 2022. The results of measurements on an AMD MI100 GPU showed that the UM version of the kernel for vector addition yielded execution times that were 1.4x to 74.8x longer than those of the non-UM version after optimization. 

While (1) the USM-based approach is not portable, as explained above, (2) the buffer-accessor approach can guarantee portability by definition. Moreover, (3) because the three cases considered above (vector addition, matrix multiplication, and reduction) show that the buffer-accessor approach yields the best performance when data migration is conducted automatically on both the CPU and the GPU, it is the best solution to ensure satisfactory performance on portable cross-platform applications.

\section{Benchmarks: NDRange vs. Hierarchical Kernel}

This section examines the performance impact of SYCL's two primary parallelism abstractions---NDRange and hierarchical kernels---across both Unified Shared Memory (USM) and buffer-accessor memory models. Our goal is to determine whether these abstractions are performance-equivalent and interchangeable, as required by the principle of singularity.

\subsection{Experimental Design}

The same testbed described in Section~5 was used: Intel's Max 1100 GPU and Xeon Platinum 8480 CPU, with SYCL kernels executed through the Intel DPC++ compiler. We measured the runtime (in milliseconds) of reduction and matrix multiplication applications implemented using both NDRange and hierarchical kernels.

Each configuration was benchmarked with both USM and buffer-accessor models to observe interaction effects. Inputs ranged from 512 MB to 2048 MB for reductions, and matrix sizes ranged from 1024 to 8192 for matrix multiplication.

\subsection{Reduction Benchmarks}

Figures~7 and~8 compare the performance of reduction kernels implemented using NDRange and hierarchical models under buffer and USM memory models.

\smallskip
\smallskip
\textbf{Findings with buffer-accessors (Figure~7):}
\begin{itemize}
    \item NDRange kernels consistently outperformed hierarchical kernels.
    \item On the GPU, NDRange was up to \textbf{2.19$\times$} faster than hierarchical.
    \item On the CPU, NDRange yielded a \textbf{1.56$\times$} speedup over hierarchical kernels.
\end{itemize}

\smallskip
\smallskip
\textbf{Findings with USM (Figure~8):}
\begin{itemize}
    \item On the GPU, NDRange remained superior, with up to \textbf{1.35$\times$} better performance.
    \item On the CPU, however, hierarchical kernels slightly outperformed NDRange by up to \textbf{1.06$\times$}.
\end{itemize}

These results indicate that the two kernel models are not interchangeable and must be selected based on target hardware, thereby violating the performance-related expectations of singularity.

\subsection{Matrix Multiplication Benchmarks}

To test a more compute-intensive application, we implemented a tiled matrix multiplication kernel using both NDRange and hierarchical models with the buffer-accessor approach. The benchmark was executed on both CPU and GPU, with OpenMP results on CPU used as a baseline.

\smallskip
\smallskip
\textbf{Results (Figure~9):}
\begin{itemize}
    \item On the GPU, NDRange massively outperformed hierarchical kernels-by \textbf{22.5$\times$ to 42.7$\times$}, depending on matrix size.
    \item On the CPU, NDRange also had a significant lead, achieving \textbf{3$\times$ to 29.2$\times$} better performance than hierarchical.
    \item The OpenMP implementation on CPU outperformed the hierarchical SYCL kernel by up to \textbf{3.1$\times$}.
\end{itemize}

These results show that hierarchical parallelism, although conceptually suited for CPUs, underperforms significantly on both target architectures.

\begin{figure*} %[!htb]
    %\captionsetup{justification=centering} %,margin=2cm}
    %\caption{Comparing results of matrix multiplication application \\on the CPU and the GPU between Buffer and USM approaches.}\label{Fig:Data1}
\begin{minipage}{.49\linewidth} 
%\begin{subfigure}{0.49\linewidth} %[h] 
         \caption{Tiled Matrix Multiplication on Intel Max 1100 GPU and Xeon 8480 CPU.}\label{Fig:Data1}
     \centering
     \includegraphics [height=7cm, width=1.0\linewidth]{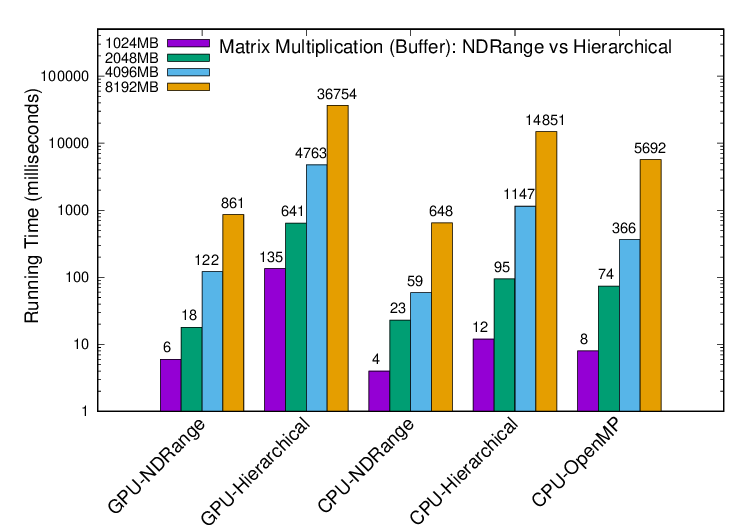}
\label{fig:fig16}  
\end{minipage} 
%\end{subfigure}
%\end{figure*} 
\hfill
%\begin{figure*}[!htb]
\begin{minipage}{.49\linewidth}
         \caption{DGEMM on NVIDIA V100 GPU and Intel Cascade Lake CPU. (From \cite{Deakin})}\label{Fig:Data1}
     \centering
     \includegraphics [height=7cm, width=1.0\linewidth]{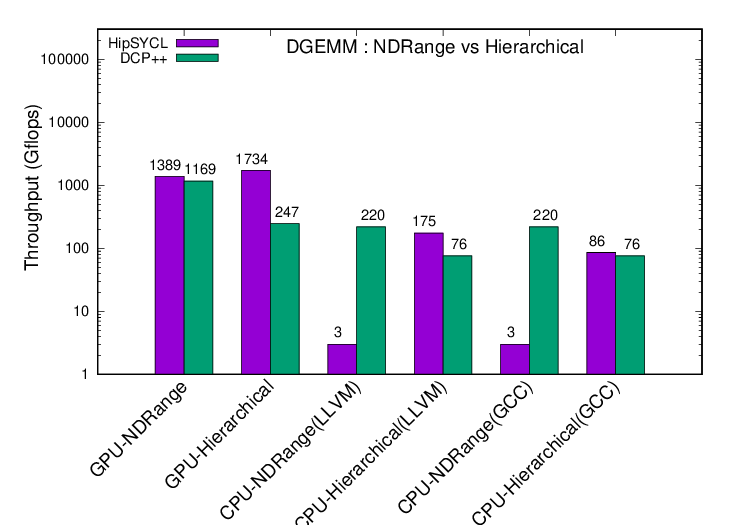}
\label{fig:fig17} 
\end{minipage}
%\end{subfigure}
\end{figure*}

\subsection{Comparison with Prior Studies}

Other researchers have reported similar variability. For example:
Deakin et al. \cite{Deakin} (2021) studied the performance of the SYCL DGEMM kernel on the Intel Cascade Lake CPU and the NVIDIA V100 GPU. They compared the performance of the DGEMM-based NDRange kernel with that of the of DGEMM-based hierarchical kernel on two SYCL implementations, hipSYCL and DCP++. They used two compilers, LLVM and GCC, on the CPU to this end. Figure 10 shows the results, in Gflops, for a matrix of size 4096 x 4096.  

It is clear from a preliminary analysis of the results that the performance profile of the DGEMM application was substantially different from that of our benchmarks. For example, the DGEMM-based hierarchical parallel kernel outperformed the DGEMM-based NDRange kernel by 25\% on the GPU when the application was implemented in hipSYCL, while the opposite result was obtained when it was implemented in DCP++, and the DGEMM-based NDRange kernel outperformed the DGEMM-based hierarchical kernel by 4.73x on the GPU. 

Another interesting observation is that regardless of the compiler used on the CPU (LLVM or GCC), the DGEMM-based NDRange kernel when implemented in DCP++ outperformed the same kernel when implemented in hipSYCL, by 73.33x. This is a difference of two orders of magnitude. Conversely, when the DGEMM-based hierarchical kernel was implemented using hipSYCL, it outperformed the same kernel when implemented by using DCP++ by up to 2.3x in case of the LLVM compiler. The clear conclusion from these results is that it is difficult to predict the performance of a SYCL application because its implementation, the compilers used, and the parallel kernel abstractions applied significantly influence its performance.

We now examine another case study from past work. Breyer
et al.\cite{Breyer} examined the scaling of the performance of the conjugate gradient (CG)-based linear equation solver at the core of their parallel least-squares support vector machine (PLSSVM) library. They compared the performance of different programing models on various architectures. We provide only the representative part of their research, including the results of the performance of the SYCL application based on two implementations (DCP++ and hipSYCL), and two parallel kernel abstractions (NDRange kernel and hierarchical kernel) on two GPUs (AMD Radeon Pro VII and NVIDIA P100), and one CPU (Intel Core-i9 10920X) for 4,096 features per data point and $2^{10}$  data points. We also provide the results of the OpenCL implementation on each platform as a reference.

\begin{figure}[!htb]
    \captionsetup{justification=centering} %,margin=2cm}
         \caption{Conjugate Gradient on NVIDIA P100 and AMD Radeon GPUs,  and Intel Corei9 10920X CPU. (From \cite{Breyer})}\label{Fig:Data1}
     \centering
     \includegraphics [height=7cm, width=1.0\linewidth]{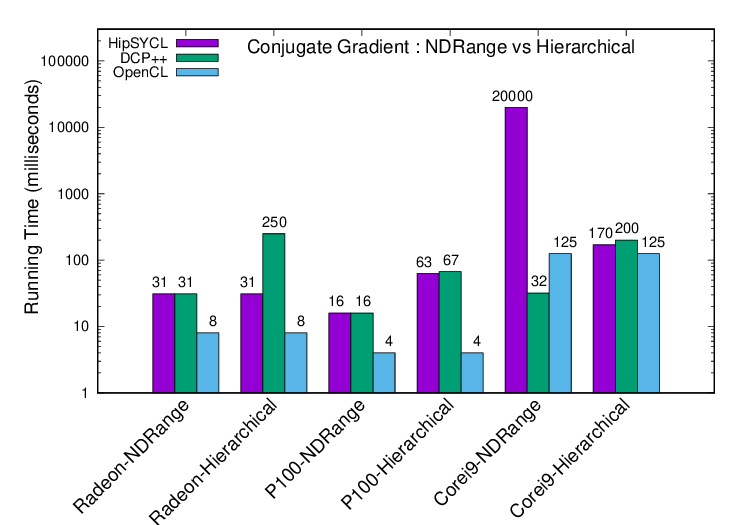}
\label{fig:fig8}  
\end{figure}

Figure 11 shows the performance (ms) of the representative platforms we chose from \cite{Breyer}.
The graphical analysis indicates unusual and unrealistic results.
For example, the NDRange kernel, when implemented by using DPC++, outperformed the same kernel when implemented by using hipSYCL on the Core-i9 CPU by 625x, which is a difference of three orders of magnitude. On the contrary, the hierarchical kernel, when using hipSYCL, outperformed the same kernel implemented by using DCP++ by 31.25x on the AMD Radeon GPU. The results obtained by OpenCL indicate ample room for improvement in the performance of the CG-based application on all platforms. For example, OpenCL outperformed SYCL by at least 16x on the NVIDIA P100 GPU.
 
Such inconsistencies demonstrate that SYCL's kernel abstractions are not only non-portable in performance, but also unpredictable without extensive benchmarking.

\subsection{Discussion}

The benchmark results and supporting literature suggest several important conclusions:

\begin{itemize}
    \item \textbf{Performance is not abstraction-agnostic.} NDRange and hierarchical kernels yield dramatically different results across platforms.
    \item \textbf{Hardware affinity is significant.} NDRange better maps to GPU architectures, while hierarchical kernels are less consistent-even on CPUs.
    \item \textbf{Compiler and implementation variation is high.} Performance can vary by orders of magnitude depending on toolchain and backend.
\end{itemize}

From a singularity standpoint, these findings are problematic. Developers must choose kernel models based on platform characteristics, breaking abstraction boundaries and undermining productivity and portability. If kernel abstractions were truly singular, developers would not have to make such architecture-specific choices.

\section{Singularity Analysis}

In this section, we evaluate how well SYCL satisfies the criteria for \textit{singularity} as defined in Section~2. We revisit the core dimensions---\textbf{portability}, \textbf{productivity}, and \textbf{performance}---and assess SYCL's capabilities based on our empirical results, illustrative examples, and literature review.

Our goal is to answer the central question: \textit{To what extent does SYCL function as a singular programming model for heterogeneous systems?}

\subsection{Portability Assessment}

SYCL offers a well-defined cross-platform abstraction layer that enables functional portability across different hardware architectures and SYCL implementations. In theory, the same SYCL source code should compile and run correctly on CPUs, GPUs, FPGAs, and other accelerators without modification.

The Unified Shared Memory (USM) model is, by definition, a software-hardware mechanism. It establishes a coherent memory space between the host and the device, enabling data to be automatically transferred between their respective memory systems. To function correctly, the USM runtime must monitor memory accesses, determine the appropriate granularity for data migration, and handle page faults---operations that can introduce significant performance overhead. Consequently, effective utilization of USM requires hardware-level support.

However, not all accelerators and GPUs currently available on the market support USM. Due to the lack of standardization in this area, even devices that claim USM support often differ in the specific features they provide. Furthermore, there is no guarantee that future devices will support USM consistently. Therefore, building a portable, cross-platform application that depends solely on USM may not be practical at present---or even in the near future. Alternative approaches should be considered.

In addition, SYCL provides two auxiliary functions, \textit{prefetch} and \textit{mem\_advise}, intended to help the USM runtime make more informed decisions about when and how to transfer data between host and device. The \textit{prefetch} function allows the application to signal when data should be proactively moved to the device, while \textit{mem\_advise} enables the programmer to specify in advance the intended usage pattern of the data (e.g., read-only or write-only access). However, these functions are not portable across platforms, as they require knowledge of the hardware's memory block size---information that is not queryable from within a SYCL application. As a result, the programmer must manually encode device-specific block size information into the application. Moreover, the naming and semantics of memory advice options differ between vendors. For example, Intel uses \textit{HW\_SPECIFIC\_ADVICE\_RO} to indicate read-only behavior, whereas AMD uses \textit{hipMemAdviseSetReadMostly} for the equivalent functionality.

Alpay et al.\ (2022)~\cite{Alpay} investigated the functional portability of SYCL by evaluating two of its major implementations: DPC++ and hipSYCL. To conduct their study, they employed all 281 benchmarks from the HeCBench \cite{Jin} benchmark suite. Their findings revealed that only 209 benchmarks (74\%) could be compiled using DPC++ without any source code modifications, while just 91 benchmarks (32\%) compiled successfully with hipSYCL in their original form.

After identifying and addressing bugs, and applying minor source code modifications, the authors were able to compile 122 benchmarks with hipSYCL---corresponding to only 43\% of the HeCBench suite. These results clearly indicate that substantial improvements are required in the functional portability of SYCL, both with respect to its specification and across different compiler implementations.

\bigskip
\bigskip

\textbf{Strengths:}
\begin{itemize}
    \item All evaluated kernel types and memory models were functionally portable across CPU and GPU.
    \item Multiple backends (OpenCL, Level Zero) and implementations (DPC++, AdaptiveCpp) support the SYCL standard to varying degrees.
\end{itemize}

\smallskip
\smallskip
\textbf{Limitations:}
\begin{itemize}
    \item Functional portability does not guarantee consistent behavior. For example, buffer lifetime, implicit data transfers, or barrier semantics may behave differently across compilers.
    \item Source-level compatibility can still be impacted by subtle implementation differences or incomplete feature support.
    \item Performance portability is not guaranteed.
\end{itemize}

\smallskip
\smallskip
\textbf{Conclusion:} \\
SYCL achieves a moderate level of \textbf{functional portability}, 
but also inconsistencies across implementations, toolchains, and runtime behaviors prevent full \textbf{portability singularity}.

\subsection{Productivity Assessment}

SYCL's single-source design and integration with modern C++ provide a high-level programming interface. However, our evaluation reveals significant barriers to developer productivity:

\smallskip
\smallskip
\textbf{Barriers to productivity:}
\begin{itemize}
    \item Developers must choose between multiple memory and kernel abstractions---each with different semantics and performance characteristics.
    \item The buffer-accessor model, while more portable, is verbose and unintuitive compared to USM.
    \item Performance inconsistencies require benchmarking and tuning even for simple workloads.
    \item Tooling support is fragmented across SYCL implementations, with differing debug, profiling, and error-reporting capabilities.
\end{itemize}

Moreover, the buffer-accessor approach is a {\it partially} implicit way to transfer data by referring to Listing 1. We have explained that once the kernel has been executed, the data updated on the device in {\it bufSum} are transferred from the device back to the host before the buffers are destroyed and the scope is freed (line 31). Note that the scope is defined by the braces on lines 6 and 31. If the programmer is not aware of how the underlying runtime behaves, and forgets to delimit the scope, {\it bufSum} is not destroyed, and no automatic migration of the data in it occurs. In this case, the data read from the total array on line 32 are not as expected. 

\smallskip
\smallskip
\textbf{Impact:} \\
These limitations impose a cognitive and engineering burden on developers, particularly those targeting multiple platforms. The need to benchmark abstraction choices (e.g., buffer vs. USM or NDRange vs. hierarchical) undermines the expected productivity benefits of a singular model.

\smallskip
\smallskip
\textbf{Conclusion:} \\
SYCL's design promotes \textbf{conceptual productivity}, but practical usage reveals considerable friction, especially for performance-critical or portable code. Thus, \textbf{productivity singularity} is only partially realized.

\begin{table*}[h]
\centering
\caption{Summary of performance characteristics for combinations of memory and parallelism abstractions evaluated in this work.}
\label{tab:perf_summary}
\resizebox{0.8\textwidth}{!}{
\begin{tabular}{l l c c c l}
\hline
\textbf{Data Model} & \textbf{Kernel Model} & \textbf{CPU Perf.} & \textbf{GPU Perf.} & \textbf{Portability} & \textbf{Notes} \\
\hline
Buffer-Accessor & NDRange      & Strong  & Strong  & Good    & Most stable across backends \\
Buffer-Accessor & Hierarchical & Moderate & Moderate & Good   & Synchronization overhead \\
USM             & NDRange      & Weak    & Mixed   & Limited & Sensitive to implicit migration \\
USM             & Hierarchical & Weak    & Weak    & Limited & Worst-case across devices \\
\hline
\end{tabular}
}
\end{table*}

\begin{table*}[h]
\centering
\caption{Alignment with singularity requirements for the evaluated abstraction combinations.}
\label{tab:singularity_summary}
\resizebox{0.5\textwidth}{!}{
\begin{tabular}{l c c c}
\hline
\textbf{Combination} & \textbf{Portability} & \textbf{Performance} & \textbf{Productivity} \\
\hline
Buffer + NDRange      & Strong  & Moderate & Moderate \\
Buffer + Hierarchical & Strong  & Weak     & Moderate \\
USM + NDRange         & Limited & Weak     & Strong   \\
USM + Hierarchical    & Limited & Weak     & Strong   \\
\hline
\end{tabular}
}
\end{table*}

\subsection{Performance Assessment}

Performance is the dimension where SYCL most clearly fails to achieve singularity. Across our benchmarks and supporting studies, we observe large disparities in runtime behavior depending on the chosen abstraction, memory model, and backend.

\smallskip
\smallskip
\textbf{Key findings:}
\begin{itemize}
    \item The buffer-accessor model consistently outperformed USM, particularly on CPU.
    \item NDRange kernels were significantly faster than hierarchical kernels on both CPU and GPU.
    \item Performance gaps exceeded \textbf{40$\times$} for matrix multiplication and \textbf{60$\times$} for reduction kernels depending on abstraction choice.
    \item SYCL compiler and implementation differences (e.g., DPC++ vs.\ hipSYCL) introduce additional variability.
\end{itemize}

\smallskip
\smallskip
\textbf{Interpretation:} \\
These results suggest that abstraction-level choices must be made based on hardware characteristics and target compilers. This violates the principle of abstraction-independence central to singularity.

\smallskip
\smallskip
\textbf{Conclusion:} \\
SYCL does not deliver consistent or predictable performance across abstraction layers or platforms. \textbf{Performance singularity} is not achieved.

\subsection{Overall Assessment of Singularity}

While SYCL brings valuable advances in single-source, cross-platform programming, it falls short of true singularity. Developers must still navigate abstraction trade-offs, tune for backend-specific behaviors, and adjust their expectations depending on target platforms.

SYCL's greatest strength is in functional unification, not performance equivalence. Until its abstractions yield more consistent outcomes across architectures, and tooling becomes more robust, \textbf{SYCL cannot be considered a fully singular programming model}.

Table~\ref{tab:perf_summary} summarizes the qualitative performance behavior of the four combinations of SYCL memory and kernel abstractions evaluated in this study. Table~\ref{tab:singularity_summary} further maps these results to the three singularity dimensions defined in Section~2: portability, performance, and productivity.

Overall, SYCL achieves reasonable portability across platforms, particularly when using buffer-accessors. However, performance remains limited, especially for hierarchical kernels and USM-based memory management. Productivity also varies depending on abstraction choice, with USM offering simpler programming but lower stability across devices.

Based on this assessment, SYCL does not yet meet full singularity. Its abstractions are not interchangeable across platforms without trade-offs in performance or portability, requiring developers to make architecture-specific decisions that singularity aims to eliminate.

\section{Discussion}

Our evaluation demonstrates that while SYCL makes significant strides toward unifying heterogeneous programming through a single-source model, it falls short of achieving full \textit{singularity} as defined by functional consistency, productivity, and performance parity across platforms. This section discusses the broader implications of our findings and outlines challenges that must be addressed to close the singularity gap.

\subsection{The Cost of Flexibility}

SYCL's design philosophy emphasizes flexibility---offering developers multiple ways to manage memory and express parallelism. This includes:

\begin{itemize}
    \item Two memory models: \textbf{Unified Shared Memory (USM)} and \textbf{buffer-accessors}
    \item Multiple kernel types: \textbf{NDRange}, \textbf{hierarchical}, and others
    \item Several backends: \textbf{OpenCL}, \textbf{Level Zero}, \textbf{CUDA}, and \textbf{HIP}
\end{itemize}

While this flexibility empowers experienced developers to tailor applications to specific hardware, it also fragments the programming experience. The lack of abstraction equivalence---where different models produce significantly different performance and behavior---undermines the principle of ``write once, run anywhere.''

In effect, the developer must act as a cross-compiler: selecting abstractions, tuning configurations, and even rewriting kernels depending on target hardware. This contradicts the goal of singularity.

Implementing hierarchical parallelism on GPUs is notoriously complex and often yields suboptimal performance. To address these challenges, the \textit{scoped parallelism} approach was proposed by Deakin et al.\ (2021)~\cite{Deakin}.

Scoped parallelism was introduced as an extension to SYCL's existing hierarchical parallelism in the \texttt{hipSYCL} implementation. Its primary goal is to enhance flexibility and performance portability by allowing the SYCL implementation greater freedom in mapping logical parallelism (as expressed by the programmer) to physical parallelism (as executed on the underlying hardware). This approach mitigates implementation challenges on both CPUs and GPUs, particularly in library-only SYCL environments, where the strict synchronization semantics of NDRange kernels can impose significant overhead.

Scoped parallelism supports nested distribution of work through the \texttt{distribute\_groups()} and \texttt{distribute\_items()} constructs, which are not limited to a fixed number of levels. All variable allocations default to private memory unless explicitly specified otherwise, thereby avoiding unexpected usage of local memory that can arise in hierarchical parallelism. Additionally, synchronization primitives such as \texttt{distribute\_items\_and\_wait()} offer fine-grained control over barriers. The scoped model is designed to be composable and implementation-agnostic, making it suitable for integration with both hierarchical and NDRange-based execution models.

Deakin et al.~\cite{Deakin} benchmarked scoped parallelism alongside NDRange and hierarchical models using SYCL-Bench micro-benchmarks (including \texttt{nbody}, \texttt{reduction}, and \texttt{segmented reduction}) as well as a tiled DGEMM (dense matrix multiplication) kernel. On an NVIDIA V100 GPU, scoped parallelism outperformed NDRange by up to 1.25x when using \texttt{hipSYCL}. Its performance was comparable to that of hierarchical parallelism and surpassed DPC++'s implementations of both NDRange and hierarchical models. These gains were attributed to better occupancy and lower register pressure. Notably, scoped parallelism introduced no performance penalty relative to NDRange abstractions in the SYCL-Bench tests.

On an Intel Cascade Lake CPU, both scoped and hierarchical models significantly outperformed NDRange for reduction workloads using \texttt{hipSYCL}, achieving speedups of up to 150x due to lower scheduling overhead. Performance differences across compilers were also observed: GCC produced better auto-vectorization than Clang for scoped parallelism. In compute-bound kernels like \texttt{nbody}, performance differences were minimal and primarily influenced by compiler vectorization quality.

Scoped parallelism appears to achieve its design goals by simplifying the programming model while improving performance portability. It combines the expressiveness of hierarchical parallelism with the runtime efficiency of NDRange on accelerators and provides a library-friendly execution model on CPUs. The study suggests that scoped parallelism is a viable---and often superior---alternative to traditional SYCL parallelism models, especially in portable high-performance computing scenarios.

While these results are promising, broader performance evaluation is needed across a wider range of platforms, compilers, and implementations before scoped parallelism can be considered for standardization within SYCL.

\subsection{Missing Semantic Guarantees}

A major obstacle to singularity in SYCL is the absence of strict semantic guarantees for abstraction behavior across platforms. Examples include:

\begin{itemize}
    \item Whether data will be implicitly copied back to the host using \texttt{set\_final\_data()} or \texttt{set\_write\_back()}
    \item Whether work-group-level synchronization is consistent across compilers
    \item How implicit memory migration behaves in USM depending on runtime
\end{itemize}

This variability forces developers to insert defensive programming constructs (e.g., host accessors to force synchronization) or conduct platform-specific testing---adding unnecessary complexity and undermining portability.

\subsection{Performance Portability as an Open Problem}

SYCL explicitly disclaims performance portability in its specification, and our results validate that decision. Performance varies widely depending on abstraction choice, memory model, and backend, with differences reaching over \textbf{40$\times$} in some cases.

In 2024, Breyer et al.~\cite{Breyer2} presented an in-depth evaluation of SYCL's various data-parallel kernel invocation methods, focusing on performance across heterogeneous hardware platforms using two major SYCL implementations: DPC++ and AdaptiveCpp.

The performance evaluation was conducted using the Parallel Least Squares Support Vector Machine (PLSSVM) library, with emphasis on kernel matrix assembly. The tested platforms included an NVIDIA A100 GPU, an AMD MI210 GPU, and a dual-socket AMD EPYC 9274F CPU.

Hyperparameters such as \texttt{THREAD\_BLOCK\_SIZE} and \texttt{INTERNAL\_BLOCK\_SIZE} were manually tuned for each 
kernel-hardware pair. The SVHN dataset, containing 73,257 RGB images, was used for benchmarking. Performance analysis was carried out using vendor-specific profilers, including NVIDIA Nsight, AMD micro-Prof, and Intel VTune.

On GPUs (NVIDIA A100 and AMD MI210), the basic data-parallel kernels consistently delivered the worst performance across all platforms. This was attributed to limited tuning capability, poor memory optimization, and high memory load overhead. In contrast, work-group kernels consistently achieved the best or near-best performance, reaching up to 71\% of the theoretical FP64 peak. These kernels demonstrated effective memory reuse and efficient use of local memory. Hierarchical and scoped parallelism in AdaptiveCpp achieved performance comparable to that of work-group kernels, particularly on the NVIDIA A100. Scoped parallelism incurred a slight overhead due to memory stalls but remained competitive on GPUs. Meanwhile, the hierarchical kernel implementations in DPC++ performed worse than work-group kernels on both the A100 and MI210, mainly due to increased instruction count and memory traffic.

On the CPU (AMD EPYC 9274F), only AdaptiveCpp was evaluated, as DPC++ relies on OpenCL, which is deprecated on AMD CPUs. The basic kernel showed poor performance due to ineffective caching and lack of memory locality. Work-group and hierarchical kernels performed similarly well, with work-group kernels holding a slight advantage. Scoped parallelism, while outperforming the basic kernel, was significantly slower due to increased wait times and poor vectorization.

In conclusion, work-group data-parallel kernels emerged as the most reliable option across platforms and compilers, offering the best overall performance and portability. Basic kernels are easy to implement but are generally unsuitable for performance-critical or cross-platform applications. Hierarchical kernels show promise, especially in AdaptiveCpp, but are hindered by incomplete specification and inconsistent compiler support. Scoped parallelism remains a promising model but is still immature on CPUs and therefore not yet recommended for portable high-performance applications.

The authors advocate for further refinement of the SYCL specification---particularly concerning hierarchical kernels---to improve performance portability and enhance usability across diverse platforms.

\begin{figure}[!htb]
    \captionsetup{justification=centering} %,margin=2cm}
         \caption{The impact of DCP++ compiler on the performance on PLSSVM library overtime (From \cite{Marcel}).}\label{Fig:Data1}
     \centering
     \includegraphics [height=7cm, width=0.9\linewidth]{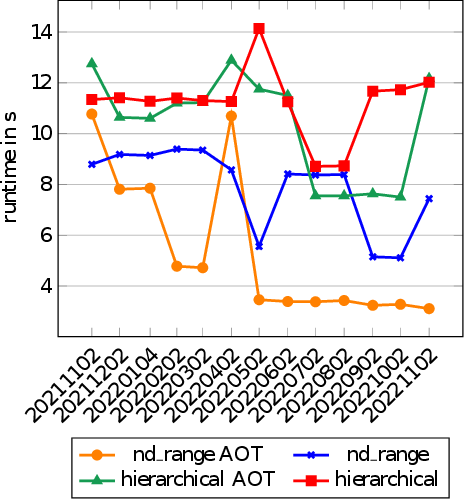}
\label{fig:fig21}  
\end{figure}

Finally, our analysis showed that the leading implementation of SYCL, the Intel DPC++ compiler, is unstable \cite{Marcel,Lin}. As an example, Figure 12 shows the results of a study by Breyer et al. \cite{Marcel} (2023) on the runtimes of a specific code of their PLSSVM library on an identical hardware, the NVIDIA A100 GPU, with identical input parameters over a year, where only the DCP++ version was changed every month. Figure 12 compares the performance of different kernel implementations, NDRange and hierarchical, on the PLSSVM library, as well as between DPC++ compilations with and without ahead-of-time compilation.  It is worth noting to the instability of the performance from one month to the next caused only by changing the versions of the compiler.

This calls into question the feasibility of using SYCL as a general-purpose programming model across architectures without architecture-specific tuning. Until compilers and runtimes mature to deliver consistent performance regardless of abstraction, developers must continue to manage performance manually.

\subsection{The Path Forward}

In the SYCL 2020 Specification, Revision 7, released on \textbf{April 18, 2023}~\cite{SYCL}, the Khronos Group stated:  
\textit{``Based on developer and implementation feedback, the hierarchical data-parallel kernel...is undergoing improvements to better align with the frameworks and patterns prevalent in modern programming.''}  
However, in Revision 10 of the SYCL 2020 Specification, released on \textbf{April 7, 2025}~\cite{SYCL2}, the Khronos Group announced that the hierarchical data-parallel kernel has been deprecated.

As a result, the SYCL developer community has lacked a clear roadmap for some time regarding the promised solutions for high-performance parallelism. The future direction of SYCL remains uncertain, and it would be highly beneficial for the SYCL designers at the Khronos Group to provide an updated and transparent roadmap outlining the standard's evolution in the coming years.

Despite the challenges described in this paper, SYCL remains a promising candidate for long-term heterogeneous programming. To move closer to singularity, the following developments are needed:

\begin{itemize}
    \item \textbf{Standardization of behavior:} Clearly defined and enforced semantics for memory, synchronization, and data movement across platforms.
    \item \textbf{Performance-aware abstractions:} New abstractions or annotations that help bridge the gap between portability and performance.
    \item \textbf{Unified tooling:} A robust, cross-implementation set of developer tools for profiling, debugging, and validating SYCL code.
    \item \textbf{Empirical tuning databases:} Compiler and runtime optimizations that learn from prior runs and adapt automatically, reducing manual tuning needs.
\end{itemize}

These changes could transform SYCL from a capable programming model into a truly singular one---fulfilling the vision of seamless cross-platform programming without sacrificing control or efficiency.

A key future challenge is to enable the compiler and runtime to automatically select architecture-appropriate memory and parallel abstractions without requiring any source-level changes, thereby moving closer to truly singular programming.

\section{Related Work}

In this section, we review a number of recent studies on SYCL performance, demonstrating that our findings and conclusions are consistent with those of previous research.

In 2017, Jarzabek and Czarnul~\cite{Czarnul} studied the performance of the CUDA Unified Memory (UM) mechanism using three benchmark applications (heat distribution, numerical integration, and Goldbach's conjecture) on two systems running CUDA 7.5. The first system was equipped with an Intel Core i5-4690K CPU and an NVIDIA GTX970 GPU, while the second had an Intel Xeon E5-2640 CPU and an NVIDIA Tesla K20m. In all cases, UM-based applications exhibited a performance degradation of 1--8\% compared to their non-UM counterparts.

In 2019, Chien et al.~\cite{Chien} evaluated two advanced CUDA memory features, \textit{memory advice} and asynchronous \textit{prefetching}, in the context of UM. Their benchmark suite, consisting of six GPU applications (Black-Scholes, matmul, conjugate gradient, Graph500, convolution, and FDTD), was tested on three systems: an Intel Core i7-7820X with GeForce GTX 1050 Ti, an Intel Xeon Gold 6132 with Tesla V100, and an IBM Power9 with NVIDIA Volta GPU (via NVLINK). Memory advice achieved performance gains of up to 34\%, while prefetching led to improvements of up to 50\% in in-memory executions.

In 2020, Gu et al.~\cite{Gu} proposed UVMBench, a unified virtual memory benchmark suite comprising 32 applications from various scientific domains. Using an Intel Xeon E5-2630 V4 CPU and an NVIDIA GTX 1080 Ti GPU with CUDA 10.2, they found an average 34.2\% slowdown for UVM applications, with maximum slowdowns reaching up to 18x for certain benchmarks such as backprop and pathfinder.

In 2021, Nozal and Bosque~\cite{Nozal} introduced Coexecutor Runtime, a scheduler for executing SYCL kernels concurrently on heterogeneous devices. Evaluated on an Intel Core i5-7500 CPU and Intel HD Graphics 630 GPU, their runtime used both USM and buffer-accessor memory models alongside static, dynamic, and HGuided load-balancing algorithms. The HGuided strategy achieved near-optimal load balancing and up to 40\% speedup (in matmul) compared to buffer-accessor. The geometric mean speedup across all benchmarks was up to 14\%.

Their results also revealed that automatic data migration in UM caused significant slowdowns. For example, explicit data movement in convolution and FDTD3d kernels on the Power9-V100 system yielded 14x and 9x speedups respectively, and up to 3x on the Core i7-7820X system. Nonetheless, UM optimizations (advice and prefetching) delivered up to 70\% and 56\% performance improvements, though explicit data movement consistently outperformed UM.

Deakin et al.~\cite{Deakin} (2021) demonstrated that SYCL performance is difficult to predict due to differences in implementations, compilers, and kernel abstractions. For instance, a DGEMM kernel using NDRange compiled with DPC++ was 73.33x slower on the CPU than the same kernel compiled with hipSYCL. Conversely, using hierarchical kernels, hipSYCL outperformed DPC++ by up to 2.3x.

Breyer et al.~\cite{Breyer} (2022) reported similar inconsistencies across SYCL implementations, with performance differences reaching up to three orders of magnitude across platforms and compilers.

In 2022, Jin and Vetter~\cite{Vetter} evaluated UM performance using AMD's HIP API and Heterogeneous Memory Management. They compared implicit UM with explicit data movement (and other methods) on an AMD MI100 GPU. Migration optimizations improved bandwidth by 48\% (vector addition, 1024K vector length), yet execution time under UM remained up to 74.8x slower than explicit copy-then-execute.

They also evaluated eight scientific benchmarks, observing consistent slowdowns in UM versions (up to 2x in GPP) compared to explicit versions. However, productivity improved, with UM versions requiring 2--24\% fewer lines of code.

Fumero~\cite{Fumero} (2022) compared USM memory types (host, shared, device) using Level Zero API on Intel GPUs. Shared and host memory outperformed device memory in a lightweight copy kernel by up to 3.5x due to reduced data migration. For the heavyweight matmul kernel, all memory types showed similar performance.

Alpay et al.~\cite{Alpay} (2022) assessed hipSYCL's USM and buffer-accessor models using the nSTREAM benchmark on NVIDIA and AMD GPUs. On the NVIDIA GTX 1080Ti, they observed up to a 10\% difference between data migration methods. On the AMD Radeon Pro VII, explicit USM performance was within 4\% of HIP and 2\% of buffer-accessor. Implicit USM, however, performed up to 100x worse.

Joube et al.~\cite{Joube} (2023) compared SYCL's USM and buffer abstractions using a PCIe-bound reduction micro-benchmark and SparseCCL clustering algorithm. On an AMD EPYC 7502 CPU and NVIDIA Quadro RTX 5000 GPU, USM caused slowdowns of 77\% and 34\%, respectively. Device-pinned USM performed similarly to buffers, but host-pinned USM performed poorly.

Mijic and Davidovic~\cite{Mijic} (2023) benchmarked SYCL DPC++ implementations of the Cholesky QR2 algorithm using both USM and buffer-accessor models. Results varied: on some CPUs and GPUs, buffer-accessor performed better; in other cases, USM did. On NVIDIA GPUs with NCCL, USM outperformed buffer-accessor by up to 50\%.

Breyer et al.~\cite{Marcel,Lin} (2023) showed that the Intel DPC++ compiler exhibited monthly runtime instability. On identical hardware (NVIDIA A100 GPU), with fixed code and inputs, performance fluctuated by up to 2x due solely to compiler version changes.

Meyer et al.~\cite{Alpay2} (2023) explored compiler-based approaches for implementing hierarchical SPMD kernels on CPUs using SYCL. Two strategies were integrated into hipSYCL: PoCL's LoopVec and Continuation-Based Synchronization (CBS).

LoopVec converts kernels by wrapping barrier-free regions in work-item loops, enabling vectorization but assuming uniform iteration counts, which limits correctness for divergent control flow. CBS partitions kernels into barrier-free sub-CFGs with a switch-based dispatcher for execution, avoiding LoopVec's uniformity assumption.

Benchmarks on Intel Ice Lake, AMD Rome, ThunderX2, and A64FX CPUs using SYCL-Bench and HeCBench revealed that CBS outperformed LoopVec on x86 systems and matched or exceeded it on ARM. CBS achieved up to 37.7x speedup over fiber-based implementations. Further CBS optimizations (e.g., loop scalarization, multidimensional nesting) yielded up to 20\% additional gains.

The authors conclude that compiler-based approaches---especially CBS---are essential for performance-portable execution of hierarchical SYCL kernels on CPUs. CBS is now the default strategy in both hipSYCL and PoCL.

\section{Conclusion}

SYCL presents a compelling vision for single-source, cross-platform programming in heterogeneous systems. By enabling developers to write unified C++ code for CPUs, GPUs, and other accelerators, it seeks to simplify application development while preserving performance and flexibility.

However, this study demonstrates that SYCL has not yet achieved \textit{singularity}---the state in which a programming model delivers portable, productive, and performant execution across platforms without requiring code modification or extensive tuning.

Our evaluation reveals three core findings:

\begin{enumerate}
    \item \textbf{Portability is partial.} While SYCL applications are functionally portable across architectures and compilers, semantic inconsistencies and backend-specific behavior often require additional care or testing.
    
    \item \textbf{Productivity is hindered by complexity.} Developers must choose between multiple memory and execution models---each with different implications for correctness, performance, and platform compatibility. This adds friction and erodes the benefits of single-source development.
    
    \item \textbf{Performance is not consistent.} Benchmark results show wide variability in runtime behavior depending on memory model, kernel type, backend, and implementation. Performance gaps exceeding \textbf{40$\times$} were observed in common workloads like reduction and matrix multiplication.
\end{enumerate}

SYCL's greatest success lies in promoting a unified programming model and accelerating the adoption of heterogeneous computing. However, to become a truly singular model, SYCL must evolve in several critical areas:

\begin{itemize}
    \item Provide stronger guarantees for abstraction behavior across compilers and runtimes
    \item Improve consistency in performance across platforms and configurations
    \item Expand tooling and diagnostic support for developers
    \item Move toward performance-aware, yet portable, abstractions
\end{itemize}

Until then, developers targeting heterogeneous systems must remain cautious: SYCL reduces the cost of portability, but not yet to zero. The dream of writing once and running efficiently everywhere remains just out of reach---but within sight.

%%%%\section*{References}

%\bibliography{mybibfile}

\end{document}